\documentclass[jap,preprint,groupedaddress,showpacs]{revtex4}

\usepackage[fleqn]{amsmath}
\usepackage{graphicx}
\usepackage{docs}
\usepackage{bm}
\usepackage[colorlinks=true,linkcolor=blue,citecolor=blue,urlcolor=black]{hyperref}%
\expandafter\ifx\csname package@font\endcsname\relax\else
 \expandafter\expandafter
 \expandafter\usepackage
 \expandafter\expandafter
 \expandafter{\csname package@font\endcsname}%
\fi

\newlength{\Figwidth}
\begin{document}

%\preprint{}

\title{Engineering Enhanced Thermoelectric Properties in 
Zigzag Graphene Nanoribbons}
\author{Hossein Karamitaheri$^{1,2}$, Neophytos Neophytou$^1$, Mahdi Pourfath$^{3,1}$, Rahim Faez$^2$, and Hans Kosina$^1$}
\affiliation{$^1$ Institute for Microelectronics, TU Wien, Wien, Austria\\
$^2$ School of Electrical Engineering, Sharif University of Technology, Tehran, Iran\\
$^3$ Electrical and Computer Engineering Department, University of Tehran,Tehran, Iran\\ ~\vspace*{1cm}
  \hspace*{0.2\linewidth} karami@iue.tuwien.ac.at \hspace{0.2\linewidth}}

\date{\today}

\begin{abstract}
We theoretically investigate the thermoelectric properties of zigzag
graphene nanoribbons in the presence of extended line defects,
substrate impurities and edge roughness along the nanoribbon's
length. A nearest-neighbor tight-binding model for the
electronic structure and a fourth nearest-neighbor force constant
model for the phonon bandstructure are used. For transport we employ quantum
mechanical non-equilibrium Green's function simulations. Starting from
the pristine zigzag nanoribbon structure that exhibits very poor
thermoelectric performance, we demonstrate how after a series of
engineering design steps the performance can be largely enhanced. Our
results could be useful in the design of highly efficient
nanostructured graphene nanoribbon based thermoelectric devices.

\end{abstract}

\pacs{72.80.Vp, 72.20.Pa, 65.80.Ck}

\maketitle
%%%%%%%%%%%%%%%%%%%%%%%%%%%%%%%%%%%%%%%%%%%%%%%%%%%%%%%%%%%%%%%%%%%%%%%%%%%%%%%%%%%%%%%
%%%%%%%%%%%%%%%%%%%% I N T R O D U C T I O N %%%%%%%%%%%%%%%%%%%%%%%%%%%%%%%%%%%%%%%%%%
%%%%%%%%%%%%%%%%%%%%%%%%%%%%%%%%%%%%%%%%%%%%%%%%%%%%%%%%%%%%%%%%%%%%%%%%%%%%%%%%%%%%%%%
\section{Introduction}
The ability of a material to convert heat into electricity is measured
by the dimensionless thermoelectric figure of merit $ZT$ defined by:
\begin{equation}
  ZT=\frac{S^2GT}{(\kappa_\mathrm{e}+\kappa_\mathrm{l})}
\end{equation}
where $S$ denotes the Seebeck coefficient, $G$ the electrical
conductance, $T$ the temperature, $\kappa_\mathrm{e}$ the electronic
and $\kappa_\mathrm{l}$ the lattice parts of the thermal
conductance~\cite{Nolas01Book}. Due to the strong interconnection
between the parameters that control $ZT$, it has been traditionally
proved difficult to achieve values above unity, which translates to
low conversion efficiencies and limit the applications for
thermoelectricity.

The recent advancements in lithography and nanofabrication, however,
have lead to the realization of breakthrough experiments on
nanostructured thermoelectric devices that demonstrated enhanced
performance, sometimes even up to 2 orders of magnitude higher than
the corresponding bulk material values. Nanostructures provided the
possibility of independently designing the quantities that control the
$ZT$ in achieving higher values. Enhanced performance was demonstrated
for 1D nanowires (NWs)~\cite{Hochbaum08,Boukai08}, 2D thin films,
1D/2D superlattices~\cite{Venkatasubramanian01,Kim06}, as well as
materials with embedded nanostructuring~\cite{Tang10}.

Graphene, a recently discovered two-dimensional form of carbon, has
received much attention over the past few years due to its excellent
electrical, optical, and thermal
properties~\cite{Novoselov04}. Graphene, however, is not a useful
thermoelectric material. Although its electrical conductance is as
high as that of copper~\cite{Chen08}, its ability to conduct heat is
even higher~\cite{Ghosh08}, which increases the denominator of
$ZT$. To make things worse, as a zero bandgap material, pristine
graphene has a very small Seebeck coefficient~\cite{Seol10}, which
minimizes the power factor $S^2G$. Nanoengineering, however, could
provide ways to increase the Seebeck coefficient and decrease the
thermal conductivity as well.

The high thermal conductivity of graphene is mostly due to the lattice
contribution, whereas the electronic contribution to the thermal
conduction is smaller~\cite{Hone99,Balandin08}. In order to reduce the
thermal conductivity, therefore, the focus is placed on reducing phonon
conduction. Recently many theoretical studies have been performed
regarding the thermal conductivity of graphene-based
structures. Several methods, such as the introduction of vacancies,
defects, isotope doping, edge roughness and boundary scattering, can
considerably reduce thermal
conductance~\cite{Jiang11,Hu10,Sevincli10}. Importantly, in certain
instances this can be achieved without significant reduction of the
electrical conductance. 

In order to improve the Seebeck coefficient graphene needs to acquire
a bandgap. This can be achieved by appropriate patterning of the
graphene sheet into nanoribbons~\cite{Han07,Zhang11}. Graphene
nanoribbons (GNRs) are thin strips of graphene, where the bandgap
depends on the chirality of the edges (armchair or zigzag) and the
width of the ribbon. Armchair GNRs (AGNRs) can be semiconductors with
a bandgap inversely proportional to their width~\cite{Han07}. Although
the acquired bandgap can increase the Seebeck coefficient, when
attempting to reduce the thermal conductivity by introducing disorder
in the nanoribbon, as described above, the electrical conductivity is
also strongly affected~\cite{Ouyang09,Areshkin07}, and the
thermoelectric performance remains low. Zigzag GNRs (ZGNRs), on the
other hand, show metallic behavior with very low Seebeck coefficient,
but as described in Ref.~\cite{Areshkin07}, the transport in ZGNRs is
nearly unaffected in the presence of line edge roughness, at least in
the first conduction plateau around their Fermi level.

In this work, by using atomistic electronic and phononic bandstructure
calculations, and quantum mechanical transport simulation, we show
that despite the zero bandgap, the thermoelectric performance of ZGNRs
can be largely enhanced. For this a series of design steps are employed:
i) Introducing extended line defects (ELDs) as described in
Ref.~\cite{Bahamon11} can break the symmetry between electrons and
holes by adding additional electronic bands. This practically provides
a sharp band edge around the Fermi level and offering a band asymmetry
which for thermoelectric purposes it practically constitutes an
``effective bandgap''. ii) Introducing background impurities enhances
the ``effective bandgap''. iii) Introducing edge roughness reduces the
lattice part of the thermal conductivity (significantly more than it
reduces the electrical conductivity). After such procedure, we
demonstrate that the figure of merit $ZT$ can be greatly enhanced and
high thermoelectric performance could be achieved.

The paper is organized as follows. In section~\ref{s:Approach} we
describe the methodology used in our calculations. In
section~\ref{s:Result} we present the results for the
electronic/phononic structure and transmission of ZGNRs for every step
of our design approach (in section~\ref{s:Transmission}), and their
influence on the thermoelectric coefficients (in
section~\ref{s:Thermoelectric}). Finally, in section~\ref{s:Summary}
we conclude.

\section{Approach}
\label{s:Approach}
In linear response regime, the transport coefficients can be evaluated
using the Landauer formula~\cite{Kim09,Jeong10,Karamitaheri11}:
\begin{equation}
G=\left (\frac{2q^2}{h}\right)I_0~~~~[1/\Omega]
\end{equation}
\vspace*{-20pt}
\begin{equation}
S=\left (-\frac{k_{\mathrm{B}}}{q}\right )\frac{I_1}{I_0}~~~~[V/K]
\label{e:seebeck}
\end{equation}
\vspace*{-20pt}
\begin{equation}
\kappa_{\mathrm{e}}=\left (\frac{2Tk_{\mathrm{B}}^2}{h}\right )\left [
  I_2-\frac{I_1^2}{I_0}\right ]~~~~[W/K]
\end{equation}
Here, $h$ is the Planck constant, $k_{\mathrm{B}}$ is the Boltzmann
constant, and
\begin{equation}
I_j=\int_{-\infty}^{+\infty}\left(
\frac{E-E_{\mathrm{F}}}{k_{\mathrm{B}}T}
\right)^j{T}_{\mathrm{el}}(E)\left(- \frac{\partial f}{\partial E} \right)dE
\end{equation}
where ${T}_{\mathrm{el}}(E)$ is the electronic transmission
probability, $f(E)$ is the Fermi function and $E_{\mathrm{F}}$ is the
Fermi-level of the system. Similarly, the lattice contribution to the
thermal conductance can be given as a function of the phonon
transmission probability~\cite{Ouyang09}:
\begin{equation}
\kappa_\mathrm{l}=\frac{1}{h}\int_{0}^{+\infty}{T}_\mathrm{ph}(\omega)\hbar\omega\left(\frac{\partial
  n(\omega)}{\partial T}\right)\ d(\hbar\omega)
\label{e:kp}
\end{equation}
where $n(\omega)$ denotes the Bose-Einstein distribution function and
${T}_\mathrm{ph}(\omega)$ is the phonon transmission probability~\cite{Jeong11}.

For the electronic structure, the Hamiltonian of the GNRs is described in the standard first nearest-neighbor atomistic tight-binding $p_z$ orbital approximation. The hopping parameter is set to $-2.7~\mathrm{eV}$ and the on site potential is shifted to zero so
that the Fermi level remains at $0~\mathrm{eV}$. This model has been
recently used to describe the electronic transport of ELD-ZGNR with
double-vacancies and the results are in good agreement with
first-principle calculations and experimental
studies~\cite{Bahamon11,Lahiri10}. To the best of our knowledge, only
a few first-principle calculations and experimental studies have been
conducted in structures that include ELDs~\cite{Appelhans10,Lusk10,Lahiri10}. The two main features of
the electronic structure, the asymmetry between electrons and holes,
and the metallic behavior of the ELD in the graphene ribbon channel
have been described in these studies, and are also captured by the
tight-binding model as we will demonstrate below.

For the phonon modes, the dynamic matrix is constructed using the
fourth nearest-neighbor force constant
model~\cite{Karamitaheri11}. The force constant method uses a set of
empirical fitting parameters and can be easily calibrated to
experimental measurements. We use the fitting parameters given in
Ref.~\cite{Saito98Book} for graphene-based structures. We assume that
this model is still valid under structures that include ELDs. Although verification of its validity for ELD-ZGNRs has not been demonstrated yet, i.e. using first-principle calculations, in
Ref.~\cite{Kahaly08} it was shown using DFT simulations that there is
little difference between the phonon transmission of carbon nanotube
structures with/without ELDs which could justify our model
choice. In any case, as we show below, the main influence on the
phonon transport in this work originates from edge roughness
scattering, which reduces the phonon transmission drastically. The
effect of edge roughness scattering is the dominant effect, and that
can be captured adequately by the model we employ in this work. The
influence of the ELDs on the phonon transmission is much smaller
compared to the effect of edge roughness, and therefore we still
choose to use the numerically less expensive fourth nearest-neighbor
force constant method.

In this work, the fully quantum mechanical non-equilibrium Green's
function formalism (NEGF) is used for transport calculations of
both electrons and phonons. The system geometry is defined as a set of two
semi-infinite contacts and a channel (device) with length $L$. The
device Green's function is obtained as
\begin{equation}
  G_{\mathrm{el}}(E)=\left (EI-H-\Sigma_{\mathrm{s},{\mathrm{el}}}-\Sigma_{\mathrm{d},{\mathrm{el}}}
  \right)^{-1}
\end{equation}
for electron calculation, where $H$ is the device Hamiltonian matrix
and $E$ is the energy. In the case of phonon transport the Green's
function is given by:
\begin{equation}
  G_{\mathrm{ph}}(E)=\left (EI-D-\Sigma_{\mathrm{s},\mathrm{ph}}-\Sigma_{\mathrm{d},\mathrm{ph}}
  \right)^{-1}
\end{equation}
where $D$ is the dynamic matrix and $E=\hbar \omega$~\cite{Karamitaheri12}. The contact
self-energy matrices $\Sigma_{\mathrm{s/d}}$ are calculated using the
Sancho-Rubio iterative scheme~\cite{Sancho85}. The effective
transmission probability through the channel can be achieved using the
relation:
\begin{equation}
  T_{\mathrm{el/ph}}(E)=\mathrm{Trace}[\Gamma_{\mathrm{s}}G\Gamma_{\mathrm{d}}G^{\dagger}]
\end{equation}
where $\Gamma_{\mathrm{s}}$ and $\Gamma_{\mathrm{d}}$ are the
broadening functions of contacts~\cite{Datta05Book}.

This method is very effective in describing the effect of realistic
distortion in nanostructures, including all quantum mechanical
effects. In our calculation, we include long-range substrate
impurities with density of one impurity per $125~\mathrm{nm}$ and edge
distortion (roughness) up to four layers in each side of the ribbon's
edge. These are applied only on the device part and not in the contact
regions~\cite{Areshkin07}.

\section{Results and Discussion}
\label{s:Result}
An efficient thermoelectric material must be able to effectively
separate hot from cold carriers. The quantity that determines the
ability to filter carriers is the Seebeck coefficient. The Seebeck
coefficient depends on the asymmetry of the density of states around
the Fermi level. In semiconductor the Seebeck coefficient is large,
but in a metal where the density of states is more uniform in energy the
Seebeck coefficient is small. Metallic ZGNRs also have a small Seebeck
coefficient because their transmission is constant around the Fermi
level, despite the peak in the DOS at $E=0~\mathrm{eV}$ due to the edge
states. Recently, however, Bahamon et al. have investigated the
electrical properties of ZGNRs that included an ELD
(ELD-ZGNRs) along the nanoribbon's length~\cite{Bahamon11}. It was
reported that the ELD breaks the electron-hole energy symmetry in
nanoribbons, and introduces an additional electron band around the
Fermi level. In such a way an asymmetry in the density of states and
the transmission function are achieved which improves the Seebeck
coefficient as we will show further down. This particular structure
has also been recently experimentally
realized~\cite{Lahiri10}. Although the method of fabrication was
rather complicated to be able to scale for industrial applications,
nevertheless it makes studies on GNRs appropriate and interesting as
well.

\subsection{Electronic and Phononic Structure}
\label{s:Transmission}
\begin{figure}[htb]
  \begin{center}
    \includegraphics[width=0.45\linewidth]{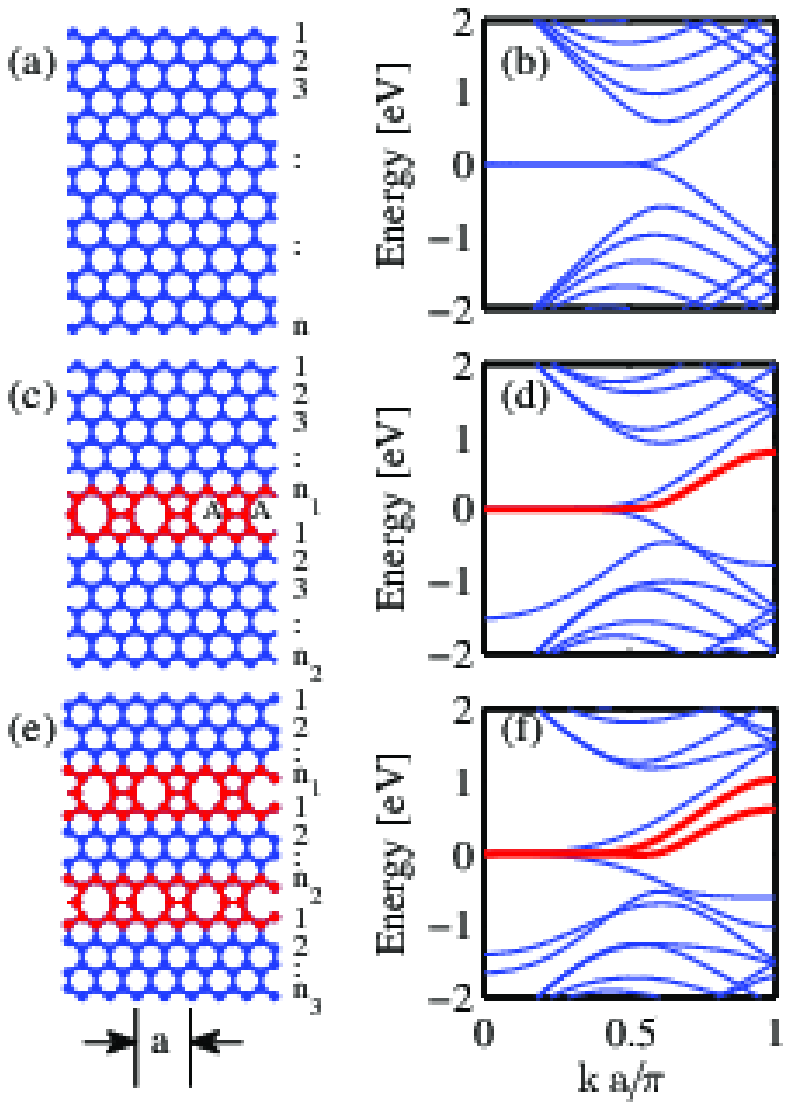}
  \caption{The geometrical structure of (a) ZGNR(n), (c)
    ELD-ZGNR($n_1$,$n_2$), and (e) 2ELD-ZGNR($n_1$,$n_2$,$n_3$). The
    bandstructure of (b) ZGNR(20), (d) ELD-ZGNR(10,10), and (f)
    2ELD-ZGNR(8,4,8). The bandstructure of ZGNR(20) is folded for a
    better comparison. The translation vector length is
    $a=0.49~\mathrm{nm}$. The $n$, $n_1$, $n_2$ and $n_3$ indicate the
    number of zigzag edges on the top, bottom, and middle of the ELD
    regions as indicated.}
\label{fig:BandStructure}
  \end{center}
\end{figure}

The changes in the electronic structure of the ZGNRs after the
introduction of the ELD are demonstrated in
Fig.~\ref{fig:BandStructure}. Figure \ref{fig:BandStructure}-a shows
the atomistic geometry of the pristine ZGNR of width $W\sim
4~\mathrm{nm}$ (with 20 zigzag edge lines) and
Fig. ~\ref{fig:BandStructure}-b its electronic structure. The Fermi
level is at $E=0~\mathrm{eV}$ due to the symmetry between electron and
hole bands. Figure ~\ref{fig:BandStructure}-c shows the structure of
the ELD-ZGNR with the same width. The region in which the ELD is
introduced is shown in red color. The ELD changes the hexagons of the
GNR to pentagons and octagons after a local rearrangement of the
bonding and the introduction of two additional atoms in the unit cell.
We use a two parameter notation to describe the ELD-ZGNR structure
throughout this work as ELD-ZGNR($n_1$,$n_2$), where $n_1$ and $n_2$
are the indices of the partial-ZGNRs above and below the line defect,
respectively (i.e. the number of zigzag edge lines of atoms), although
in all cases we use $n_1=n_2$. The bandstructure of the
ELD-ZGNR(10,10) is shown in Fig.~\ref{fig:BandStructure}-d. The
thick-red line shows a new band that is introduced in the conduction
band near the Fermi energy ($E=0~\mathrm{eV}$), which corresponds to
the ELD. There are two points that result in the creation of the extra
band. Part of the physics behind this is explained by Pereira et
al. in Ref.~\cite{Pereira06}. The first point is that a defect in the
graphene system will introduce states that reside close to the Fermi
level at $E=0~\mathrm{eV}$. This is similar to the edge states of the
ribbons that tend to reside near the Fermi level. The second point
again described in Ref.~\cite{Pereira06}, is that an asymmetry in the
dispersion between electrons and holes will be created when carbon
atoms of the graphene sublattice ``A'' (or ``B'') are coupled with
atoms from ``A'' (or ``B'') again. Usually, the atomic arrangement in
graphene can be splitted in sublattices ``A'' and ``B'', where atoms
from ``A'' couple to ``B'' and vise versa. When this happens, the
dispersion is symmetric in the first-nearest neighbor tight-binding model. At a
defect side such as the ELD we consider, where ``A'' connects to ``A''
as seen in Fig.~\ref{fig:BandStructure}-c, such asymmetry can be
observed. The fact that the overall bandstructure has additional bands
compared to the pristine ribbon is also connected to the two extra
atoms in the unit cell.

Moving one step
further, in Fig.~\ref{fig:BandStructure}-e we show the geometry of a
GNR with two ELDs. We denote this structure as
2ELD-ZGNR($n_1$,$n_2$,$n_3$), where $n_1$, $n_2$, and $n_3$ denote the
the number of zigzag carbon lines above, within, and below the line
defects. Figure~\ref{fig:BandStructure}-f shows the electronic
structure of the 2ELD-ZGNR(8,4,8). In this case two additional bands
are introduced near the Fermi level as noted by the thick-red
lines. In this structure the asymmetry between electron and hole bands
around the Fermi level ($E=0~\mathrm{eV}$) is further enhanced.

\begin{figure}[htb]
  \begin{center}
    \includegraphics[width=0.45\linewidth]{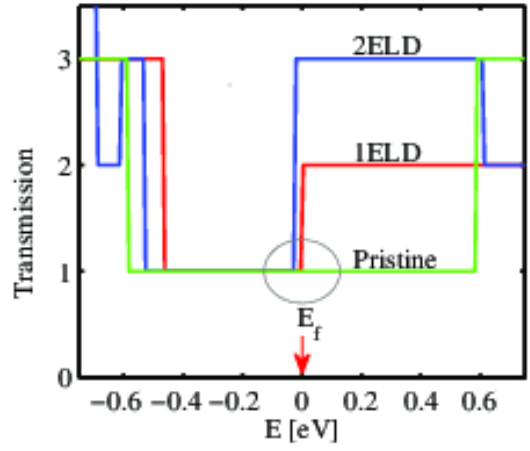}
  \caption{The transmission function for three different structures:
    i) The pristine ZGNR(20), ii) ELD-ZGNR(10,10), and iii)
    2ELD-ZGNR(8,4,8).}
\label{fig:PristineTrans}
  \end{center}
\end{figure}

$\mathbf{1^{st}}$ {\bf Design Parameter- The Effect of ELD:} Figure
\ref{fig:PristineTrans} demonstrates the increase in the asymmetry of
the bands around the Fermi level by showing how the transmission
changes when one or two ELDs are introduced in the
channel. For the pristine ZGNR, the transmission is equal to one,
indicating the existence of a single propagating band at energies
around the Fermi level (green line). With the introduction of one ELD,
the conduction band ($E>0~\mathrm{eV}$) is composed of two subbands,
whereas the valence band ($E<0~\mathrm{eV}$) is still composed of one
subband. With the introduction of two ELDs, three conduction subbands
now appear, but still only one valence subband. As it
will be shown below, this asymmetry will improve the Seebeck
coefficient. This constitutes the first design step in improving the
thermoelectric performance of ZGNRs.

\begin{figure}[htb]
  \begin{center}
    \includegraphics[width=0.45\linewidth]{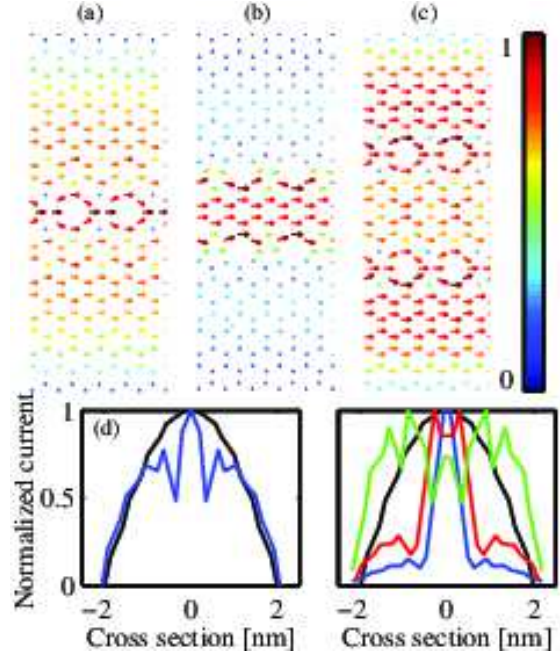}
  \caption{Normalized current spectrum at $E=0.2~\mathrm{eV}$ for (a)
    ELD-ZGNR(10,10), (b) 2ELD-ZGNR(8,4,8), and (c) 2ELD(7,6,7). (d)
    The current in the cross section of ZGNR(20) (black line)
    and ELD-ZGNR(10,10) (blue line). (e) The current in the
    cross section of ZGNR(20) (black line), 2ELD-ZGNR(9,2,9)
    (blue line), 2ELD-ZGNR(8,4,8) (red line), and
    2ELD-ZGNR(7,6,7) (green line).}
\label{fig:Current}
  \end{center}
\end{figure}

There is, however, another point worth mentioning. In
Fig.~\ref{fig:Current} we show colormaps of the normalized current
spectrum at $E=0.2~\mathrm{eV}$ in the cross sections of the ELD-ZGNRs described in Fig.~\ref{fig:PristineTrans}. Figure
~\ref{fig:Current}-a shows the current spectrum of the
ELD-ZGNR(10,10). The current is zero close to the edges of the ribbon
and peaks near the center. This is demonstrated more clearly in
Fig. \ref{fig:Current}-d, which shows the current along one atomic
chain perpendicular to this channel (blue line). The black line of
Fig.~\ref{fig:Current}-d illustrates the current density on the cross
section of the pristine ZGNR channel for reference.

The current spectrum for the 2ELD-ZGNR(8,4,8) is shown in
Fig.~\ref{fig:Current}-b. The situation is now different since most of
the current is confined within the two ELDs. This, however, is the
case only when the distance between the ELDs is smaller than the
widths of the upper/lower regions. In the case where the width of the
middle region similar to the widths of the upper/lower
regions, the current is spread more uniformly in the channel as shown
in Fig.~\ref{fig:Current}-c for the 2ELD-ZGNR(7,6,7)
channel. Figure~\ref{fig:Current}-e shows again the current along one
atomic chain in the cross section of these ribbons. The current spectrum
is localized in the middle of the channel in the 2ELD-ZGNR(8,4,8)
channel (red line) compared to the pristine channel (black line). In a
2ELD-ZGNR(9,2,9) channel with a narrower middle region the current
spectrum is localized even closer around the center (blue line). A
large portion of the current is in general flowing around the ELD
regions. The design capability to localize the current spectrum in the
middle of the channel away from the edges will prove advantageous in
the presence of edge roughness since the current in this case will be
less affected. On the other hand, in the case of the 2ELD-ZGNR(7,6,7)
channel the current spectrum tends to concentrate more close to the
edges (green line).

\begin{figure}[htb]
  \begin{center}
    \includegraphics[width=0.45\linewidth]{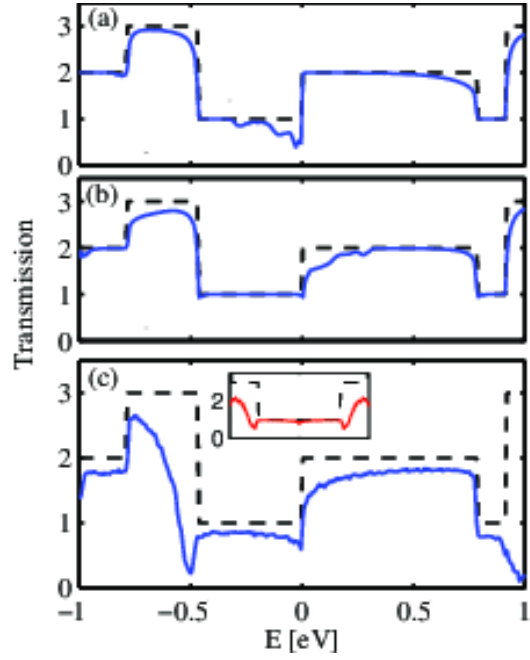}
  \caption{The effect of (a) positive substrate impurity, (b) negative
    substrate impurity, and (c) roughness on the transmission of
    ELD-ZGNR(10,10) with length of $125~\mathrm{nm}$. Inset of (c):
    The transmission of ZGNR(20) in the presence of roughness.}
\label{fig:RoughImpurity}
  \end{center}
\end{figure}

$\mathbf{2^{nd}}$ {\bf Design Parameter- The Effect of Background
  Positive Impurities:} We next illustrate the possibility of further
enhancing the asymmetry between electron and hole transport near the
Fermi level by the introduction of positively charged substrate background
impurities. The effect of background impurities is included in the
Hamiltonian in a simplified way as an effective negative long range
potential energy on the appropriate on-site Hamiltonian elements as
described in Ref.~\cite{Areshkin07}. A positive impurity in the
substrate will constitute a repulsive potential for holes (a barrier
for holes but a well for electrons) and will degrade hole transport
more effectively than electron
transport. Figure.~\ref{fig:RoughImpurity}-a shows how the
transmission of the ELD-ZGNR(10,10) channel (dashed-black line) is
affected after the introduction of positive charged impurities in the
channel (solid-blue line). Indeed, the transmission of holes below the
Fermi level ($E=0~\mathrm{eV}$) is degraded. This effect additionally
increases the asymmetry of the propagating bands and improves the Seebeck
coefficient. On the other hand, the opposite is observed when negative
impurities are introduced in the substrate. Negative impurities are a
barrier for electrons and reduce their
transmission~\cite{Neophytou06}, but do not interfere with the hole
subsystem as shown in Fig.~\ref{fig:RoughImpurity}-b. This type of
impurities will actually harm the asymmetry and needs to be avoided.

$\mathbf{3^{rd}}$ {\bf Design Parameter- The Effect of Roughness:} In
the third step of the design process we introduce the effect of edge
roughness. The inset of Fig.~\ref{fig:RoughImpurity}-c shows the
influence of edge roughness on the transmission of the ZGNR(20) of
length $125~\mathrm{nm}$. As also described in previous
studies~\cite{Areshkin07,Sevincli10}, in the first conduction plateau
the effect is negligible. In contrast to ZGNR, ELD-ZGNRs as well as
2ELD-ZGNRs are affected by edge roughness. This is because the
bandstructure of these GNRs has undergone a band folding, and
therefore, the states in the first conduction plateau have lower wave
vectors. As the long range defects can induce only small value of
momentum transfer, the momentum conservation rule indicates that, in
contrast to the ZGNR, the transport of ELD-ZGNRs and 2ELD-ZGNRs will
not remain ballistic in the presence of line edge roughness and long
range substrate impurities. This is shown in
Fig.~\ref{fig:RoughImpurity}-c, where the transmission of a roughened
$125~\mathrm{nm}$ long ELD-ZGNR(10,10) channel (solid-blue line) is reduced by
$\sim 25\%$ compared to the ballistic value (dashed-black line). Edge
roughness degrades the conductivity of holes and electrons by a
similar amount, and therefore, the level of asymmetry around the Fermi
level is retained.

\begin{figure}[htb]
  \begin{center}
    \includegraphics[width=0.45\linewidth]{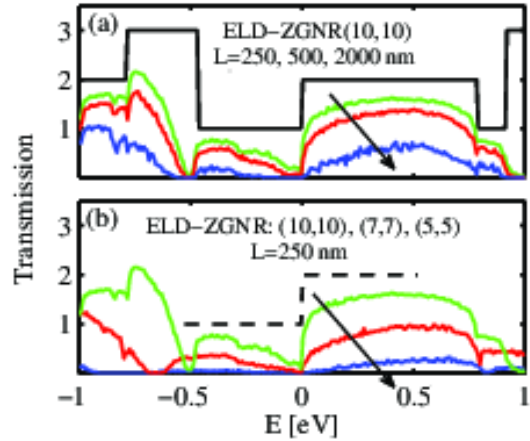}
  \caption{The influence of roughness and positive impurities on the
    ELD-ZGNR channel. (a) Electronic transmission of
    ELD-ZGNR(10,10). Rough edges are assumed and the length L is
    varied. The arrow indicates increasing values of length $L$. (b)
    Electronic transmission of ELD-ZGNRs with different widths. The
    length is assumed to be constant at $250~\mathrm{nm}$ and the
    arrow indicates the direction of decreasing the ribbon's
    width. Black-solid and black-dashed lines in (a) and (b): The
    transmission of the pristine ELD-ZGNR.}
\label{fig:ElTrans}
  \end{center}
\end{figure}

Figures \ref{fig:ElTrans}-a and \ref{fig:ElTrans}-b illustrate the
influence of roughness in ELD-ZGNR channels on their transmission, for channels of different lengths and widths. In this calculation positive impurities are also
included. Figure \ref{fig:ElTrans}-a shows the transmission of edge
roughened ELD-ZGNR(10,10) versus energy for the channel lengths
$L=250$, $500$, and $2000~\mathrm{nm}$. As the channel length is increased, the
transmission drops further compared to the transmission of the ideal
channel (black-solid line). This is expected since the channel
resistance increases with increasing length. Figure
\ref{fig:ElTrans}-b illustrates the effect of the ribbon's width on the
transmission of ELD-ZGNRs with rough edges. In this case the length is
kept constant at $L=250~\mathrm{nm}$, and results for three different ribbon
with parameters (10,10), (7,7), and (5,5) are shown. As the width of
the ribbon is decreased, the effect of line edge roughness scattering
on the transmission becomes stronger because the carriers reside on
average closer to the edges.

It is worth mentioning that the effect of edge roughness on the
transmission is much stronger in AGNR than in ZGNR. Although in the
case of some AGNRs a bandgap is naturally present and the asymmetry
does not need to be created with the introduction of line defects and
impurities, the conductance is severely degraded by the roughness
which renders this type of ribbon not well suited for transport
applications~\cite{Areshkin07}. (Note that edge roughness will be
needed in order to reduce thermal conductivity as will be shown
below.)

\begin{figure}[htb]
  \begin{center}
    \includegraphics[width=0.45\linewidth]{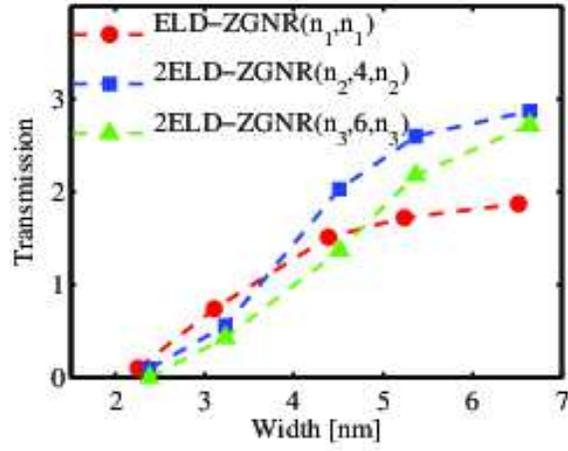}
  \caption{Transmission at $E=0.2~\mathrm{eV}$ for three different
    structures as indicated versus their width. The length is assumed
    to be constant at $250~\mathrm{nm}$.}
\label{fig:Width}
  \end{center}
\end{figure}

As we mentioned above in Fig.~\ref{fig:Current}, the channel which
includes two ELDs can shift the majority of the
current spectrum in the region between the two ELDs, and thus farther
away from the edges. It is therefore expected that the 2ELD-ZGNR will
be less affected by edge roughness scattering than the ELD-ZGNR. A
comparison of the transmission of these devices with rough edges is
shown in Fig.~\ref{fig:Width}. The transmission of
ELD-ZGNR($n_1$,$n_1$), and two cases of 2ELD-ZGNR,
2ELD-ZGNR($n_2$,4,$n_2$) and the 2ELD-ZGNR($n_3$,6,$n_3$) at
$E=0.2~\mathrm{eV}$ versus their width $W$ are compared. The
parameters $n_i$ are adjusted such that the three channels have nearly
the same width $W$. The first channel belongs to the category shown in
Fig.~\ref{fig:Current}-a, the second in the category of
Fig.~\ref{fig:Current}-b, and the third in the category of
Fig.~\ref{fig:Current}-c. The third channel as shown in
Fig.~\ref{fig:Current} spreads the current spectrum more uniformly in
the channel and is expected to be affected the most from edge
roughness. All channels have the same length of $L=250~\mathrm{nm}$. For
smaller widths the effect of roughness is strong, and the
transmissions of all channels are drastically reduced. Since the
2ELD-ZGNR devices can concentrate the current spectrum around the
defect lines as shown in Fig.~\ref{fig:Current}-b and
\ref{fig:Current}-c, they effectively bring it closer to the edges and
the reduction is larger for these devices. For larger widths the
transmission of the ribbons approaches its ballistic value, which is 2
for the ELD-ZGNR devices and 3 for the 2ELD-ZGNR devices. The
transmission of the 2ELD-ZGNR($n_2$,4,$n_2$) channels increases
faster with increasing channel width, because the current spectrum is located farther from the
edges which makes it less susceptible to scattering as the width
increases. The transmission of 2ELD-ZGNR($n_3$,6,$n_3$) channel
eventually increases close to the ballistic transmission value as the
width increases, but it increases more slowly than that of the
2ELD-ZGNR($n_2$,4,$n_2$) channel.

\begin{figure}[htb]
  \begin{center}
    \includegraphics[width=0.45\linewidth]{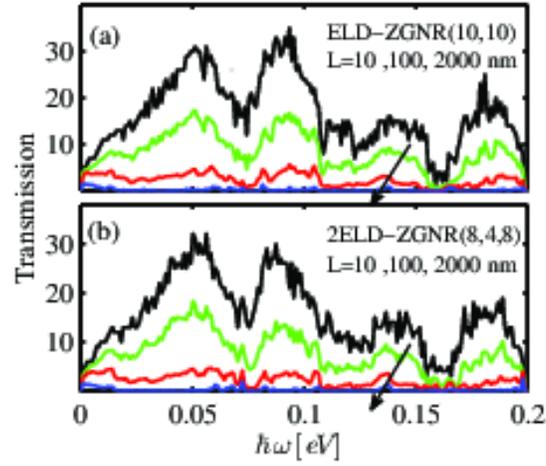}
  \caption{Phonon transmission probability of (a) ELD-ZGNR(10,10) and
    (b) 2ELD-ZGNR(8,4,8). Rough edges are assumed and the length $L$
    of the channel is varied. The arrows indicate increasing values of
    channel length $L$, $10~\mathrm{nm}$-green, $100~\mathrm{nm}$-red,
    $2000~\mathrm{nm}$-blue lines. Black lines: The phonon
    transmission of the channels with line defects but without
    roughness.}
\label{fig:PhTrans}
  \end{center}
\end{figure}

{\bf Effect of roughness on phonon Transport:} Although the reduction
in the electronic transmission of channels with ELDs can be quite
strong when considering edge roughness, the reduction in the lattice
part of the thermal conductivity is even stronger. We take advantage of on this
effect when attempting to optimize the thermoelectric figure of
merit. The phonon transmission for the edge roughened ELD-ZGNR(10,10)
channel versus energy is shown in Fig.~\ref{fig:PhTrans}-a. Results
for channel lengths $L=10$, $100$, and $2000~\mathrm{nm}$ are shown. As
expected, the transmission decreases as the length is increased. What
is important, however, is that the decrease is much stronger than the
decrease of the electron transmission shown in
Fig.~\ref{fig:ElTrans}-a. For example, for a channel length of
$L=100~\mathrm{nm}$ the phonon transmission reduces by more than a factor of
$6X$, whereas the electronic transmission even at larger length
$L=250~\mathrm{nm}$ reduces only by $<30\%$. Interestingly, the same
order of reduction of the phonon transmission is observed for the
2ELD-ZGNRs as shown in Fig.~\ref{fig:PhTrans}-b, indicating that the
line defect does not affect phonon conduction significantly compared to the effect of edge roughness.

\subsection{Thermoelectric Coefficients}
\label{s:Thermoelectric}
\begin{figure}[htb]
  \begin{center}
    \includegraphics[width=0.45\linewidth]{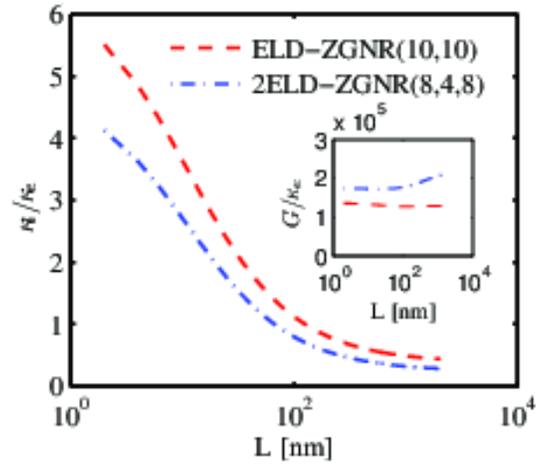}
  \caption{The ratio of the phononic to the electronic thermal
    conductivity versus cannel length $L$ for the ELD and 2ELD
    structures as noted. Inset: The ratio of the electronic
    conductivity to the electronic part of the thermal conductivity
    versus channel length $L$.}
\label{fig:NormalizedThermal}
  \end{center}
\end{figure}

The denominator of the $ZT$ figure of merit consists of the summation
of the contributions to the thermal conductivity of the electronic
system and the phononic system. In graphene the phonon part dominates
the thermal conductivity, whereas the electronic part contribution is
much smaller. The situation is different,however, in rough ELD-ZGNRs,
in which the phonon thermal conductivity is degraded more than the
electronic thermal conductivity. Figure \ref{fig:NormalizedThermal}
clearly illustrates this effect by showing the ratio of the phonon
thermal conductance to the electronic thermal conductance versus the
rough channel length. The cases of ELD-ZGNR(10,10) and
2ELD-ZGNR(8,4,8) are shown in dashed-red and dash-dot-blue lines,
respectively. For small channel lengths, where transport is
quasi-ballistic and roughness does not affect the transmission
significantly, $\kappa_{\mathrm{l}}$ is almost $5X$ larger than
$\kappa_{\mathrm{e}}$. As the length of the channel increases and the
effect of the roughness becomes significant, the phonon system is
degraded more than the electronic system, and the
$\kappa_{\mathrm{l}}$ is significantly reduced compared to
$\kappa_{\mathrm{e}}$. For lengths $L\sim 100~\mathrm{nm}$ and beyond,
$\kappa_{\mathrm{l}}$ can become even smaller than
$\kappa_{\mathrm{e}}$. The trend is the same when considering channels
with one or two ELDs. We note that from the inset of
Fig.~\ref{fig:NormalizedThermal} which shows that the ratio of the electrical
conductance $G$ over $\kappa_{\mathrm{e}}$ is almost constant, it can be
indicated that both $G$ and $\kappa_{\mathrm{e}}$ follow the same
trend, as the Wiedemann-Franz law dictates. We mention that the
$\kappa_{\mathrm{l}}$ and $\kappa_{\mathrm{e}}$ values used in
Fig.~\ref{fig:PhTrans} are extracted using the corresponding mean free
paths (MFPs) for phonons and electrons respectively, defined as
described in Ref.~\cite{Sevincli10}
\begin{equation}
T(E)=\frac{N_{\mathrm{ch}}(E)}{1+\frac{L}{\lambda(E)}}
\end{equation}
where, $T(E)$ is transmission probability, $N_{\mathrm{ch}}(E)$ is the
number of modes at energy $E$, $L$ is the given length of the channel,
and $\lambda(E)$ is the mean free path of the carriers. Alternatively,
$\kappa_{\mathrm{l}}$ and $\kappa_{\mathrm{e}}$ could be extracted
from the transmission calculations by using a statistical average over
several rough samples for each channel length. The results of both
methodologies are in good agreement for the electronic part of the
thermal conductivity. For the lattice part, the agreement is good only
for the shorter channels, below $\sim 100~\mathrm{nm}$. For larger channel
lengths, the phonon transmission is severely reduced which increases
the noise in the calculation for extracting the
$\kappa_{\mathrm{l}}$. The values extracted directly from the
integration of the phonon transmission could be as much as $2X$
larger, which could increase the
$\kappa_{\mathrm{l}}/\kappa_{\mathrm{e}}$ by a factor of $2X$ for the
longer channels. In this case the ratio
$\kappa_{\mathrm{l}}/\kappa_{\mathrm{e}}$ will be closer to unity, but
this is still a huge advantage compared to devices without roughness.

\begin{figure}[htb]
  \begin{center}
    \includegraphics[width=0.40\linewidth]{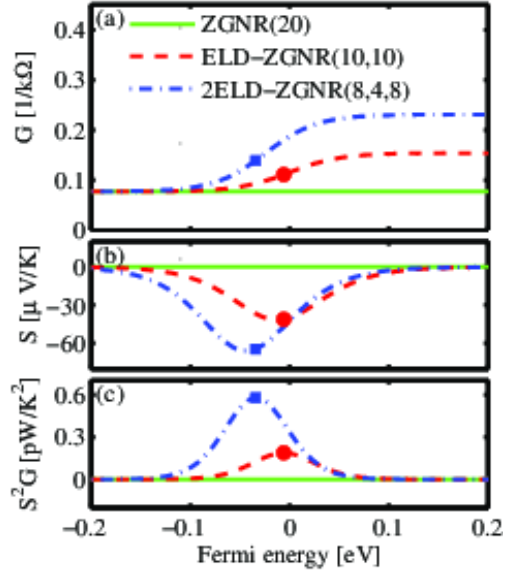}
  \caption{(a) Electrical conductance, (b) Seebeck coefficient, and
    (c) thermoelectric power factor of pristine ZGNR(20),
    ELD-ZGNR(10,10), and 2ELD-ZGNR(8,4,8) channels with perfect edges
    . The dots indicate the Fermi energy values at which the peak of
    the power factor occurs for the ELD and 2ELD channels.}
\label{fig:ThermoelectricP}
  \end{center}
\end{figure}

{\bf Power Factor:} Using the first design step, i.e. the effect of
ELDs, we have demonstrated that the transmission of electrons around
the Fermi level can be increased (from $T=1$ to $T=2$ and $T=3$ in the
presence of one and two ELDs, respectively). An asymmetry is thus
created between holes and electrons. This increases both the
conductivity and Seebeck coefficient of the channel as shown in
Fig.~\ref{fig:ThermoelectricP}. Figure \ref{fig:ThermoelectricP}-a
shows the conductance of the 2ELD-ZGNR(8,4,8) (blue), of the ELD-ZGNR
(10,10) (red), and of the pristine nanoribbon (green) at room
temperature $300~K$. As expected, the conductance of the channel with
two ELDs is the largest, followed by the channel with one ELD. They
are larger than the pristine channel by $\sim 3X$ and $\sim 2X$,
respectively. Figure \ref{fig:ThermoelectricP}-b shows the changes of
the Seebeck coefficient after the introduction of the ELDs in the
nanoribbon. Due to its metallic behavior and the flat transmission near the Fermi level, the pristine channel exhibits zero
Seebeck coefficient. Due to the built asymmetry after the introduction
of the ELDs, however, the Seebeck coefficient increases for both
channels. The channel with two line defects has the largest asymmetry,
and therefore the largest Seebeck coefficient (in absolute
values). Finally, the power factor in Fig.~\ref{fig:ThermoelectricP}-c
is indeed largely improved in the ELD structures, and
especially the 2ELD-ZGNR channel.

\begin{figure}[htb]
  \begin{center}
    \includegraphics[width=0.40\linewidth]{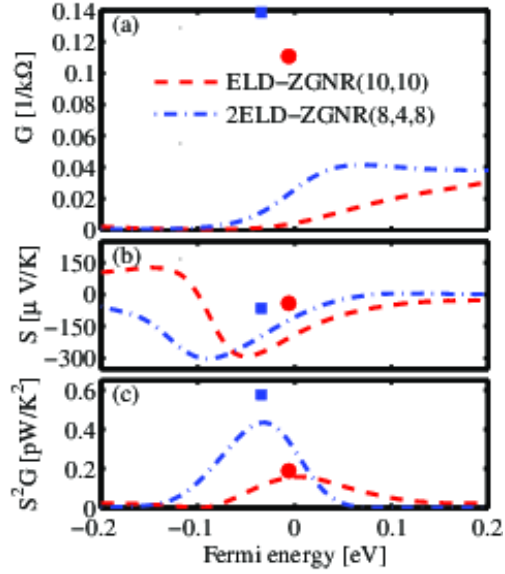}
  \caption{(a) Electrical conductance, (b) Seebeck coefficient, and
    (c) thermoelectric power factor of ELD-ZGNR(10,10) and
    2ELD-ZGNR(8,4,8) with rough edges and positively charged substrate
    impurities. The channel length is $2~\mu m$. The dots indicate the
    Fermi energy values at which the peak of the power factor occurs
    for the pristine ELD and 2ELD channels of
    Fig.~\ref{fig:ThermoelectricP} for comparison purposes.}
\label{fig:ThermoelectricL}
  \end{center}
\end{figure}

In Figure \ref{fig:ThermoelectricL} we show the same thermoelectric
coefficients for the same structures as in
Fig.~\ref{fig:ThermoelectricP}, but now edge roughness and positive
impurities are included in the calculation. The length of the channels
in this case is $2000~\mathrm{nm}$. A similar qualitative behavior is observed
as in Fig.~\ref{fig:ThermoelectricP} for both
channels. Quantitatively, however, the conductance in
Fig.~\ref{fig:ThermoelectricL}-a is now significantly reduced by a
factor of $\sim 15X$ (the dots correspond to the position of the peak
of the power factor of the devices without roughness and impurities in
Fig.~\ref{fig:ThermoelectricP}). The Seebeck coefficient in
Fig.~\ref{fig:ThermoelectricL}-b, on the other hand
increases. Finally, the peak of the power factor in
Fig.~\ref{fig:ThermoelectricL}-c reduces only slightly compared to the
peak of the power factor of the devices without edge roughness in
Fig.~\ref{fig:ThermoelectricP}-c (dots).

\begin{figure}[htb]
  \begin{center}
    \includegraphics[width=0.45\linewidth]{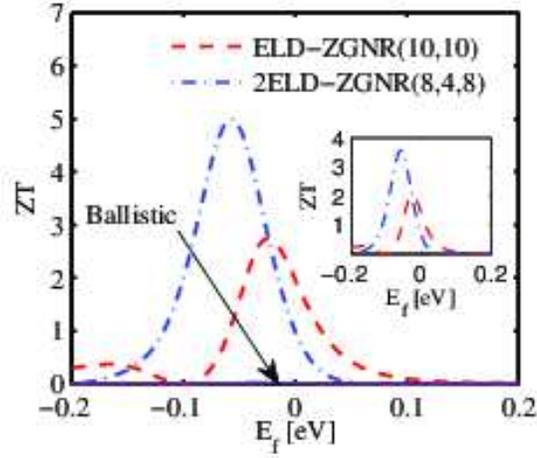}
  \caption{The thermoelectric figure of merit $ZT$ for the
    ELD-ZGNR(10,10) (dashed-red line) and 2ELD-ZGNR(8,4,8)
    (dash-dot-blue line) channels of length $L=2~\mu m$. The lattice
    thermal conductance is extracted from the calculated mean free
    path. Inset: The same figure of merit $ZT$ but with the lattice
    thermal conductance extracted by integrating the simulated phonon
    transmission.}
\label{fig:ZT}
  \end{center}
\end{figure}

{\bf Thermoelectric Figure of Merit:} For the devices that include
rough edges, however, as we demonstrated in
Fig.~\ref{fig:NormalizedThermal}, the phonon thermal conductivity is
drastically reduced compared to the electronic thermal conductivity. A
large improvement is therefore expected in the $ZT$ figure of
merit. Figure \ref{fig:ZT} shows the $ZT$ figure of merit versus
energy at room temperature for the ELD-ZGNR(10,10), the
ELD-ZGNR(10,10) with impurities and roughness (red), and the
2ELD-ZGNR(8,4,8) (blue) with impurities and roughness. As indicated,
large values of $ZT$ can be achieved, especially in the case of the
device with two ELDs. The phonon lattice conductivity value used in
this calculation was extracted using the MFP method. Since as
explained above, that value could be $2X$ lower than the value
extracted from direct integration of the 'noisy' transmission, in the
inset of Fig.~\ref{fig:ZT} we show the $ZT$ versus energy using the
$\kappa_{\mathrm{l}}$ values extracted from the transmission. Indeed
the values could be reduced by a factor of $\sim 2X$, but still peak
$ZT$ values above 2 can be achieved at room temperature, which is
comparable and even better than the best thermoelectric materials to
date~\cite{Snyder08}. We note that as shown by Ref.~\cite{Sevincli10}
rough ZGNRs can have high $ZT$ values even without the presence of
ELDs. For this however, the asymmetry in the sharp edges of the higher
subbands is utilized at energies above $0.5~\mathrm{eV}$. Those
energies however, are too high and can not easily be reached. Finally
we mention here that our formalism has considered scattering only by
edge roughness and impurity scattering, whereas phonon scattering and
dephasing mechanisms are not included. However, as it is shown for 1D
NWs~\cite{Neophytou11}, the effects of impurity scattering and edge
roughness are the most important scattering effects in channels of
cross sections below $5~\mathrm{nm}$, and we expect this to hold also for GNRs
as well.

\section{Summary}
\label{s:Summary}
In this work we present a theoretical design procedure for achieving
high thermoelectric performance in zigzag graphene nanoribbon (ZGNRs)
channels, which in their pristine form have very poor performance. The
fully quantum mechanical non-equilibrium Green's function technique was
used for electron and phonon transport, and tight-binding and force
constant methods were used for the electronic and phonon bandstructure
descriptions. We show that by introducing extended line defects (ELDs)
in the length of the nanoribbon we can create an asymmetry in the
density of modes around the Fermi level, which improves the Seebeck
coefficient. ELDs increase the electronic conduction modes, which
increase the channel conductance as well. The power factor is
therefore significantly increased. In addition, we show that by
introducing edge roughness the phonon thermal conductivity
($\kappa_{\mathrm{l}}$) is drastically degraded much more than the
electronic thermal conductivity ($\kappa_{\mathrm{e}}$), or the
electronic conductance ($G$). These three effects result in large
values of the thermoelectric figure of merit, and indicate that
roughed ZGNRs with ELDs could potentially be used as efficient high
performance thermoelectric materials.  

%%%%%%%%%%%%%%%%%%%%%%%%%%%%%%%%%%%%%%%%%%%%%%%%%%%%%%%%%%%%%%%%%%%%%%%%%%%%%%%%%%%%%%%
%%%%%%%%%%%%%% A C K N O W L E D G M E N T S  %%%%%%%%%%%%%%%%%%%%%%%%%%%%%%%%%%%%%%%%%
%%%%%%%%%%%%%%%%%%%%%%%%%%%%%%%%%%%%%%%%%%%%%%%%%%%%%%%%%%%%%%%%%%%%%%%%%%%%%%%%%%%%%%%
\section*{Acknowledgments}
This work, as part of the ESF EUROCORES program EuroGRAPHENE, was
partly supported by funds from FWF, contract I420-N16. This work was
also partly supported by the Austrian Climate and Energy Fund, contract
No. 825467.
%%%%%%%%%%%%%%%%%%%%%%%%%%%%%%%%%%%%%%%%%%%%%%%%%%%%%%%%%%%%%%%%%%%%%%%%%%%%%%%%%%%%%%%
%%%%%%%%%%%%%%%%%%%% R E F E R E N C E S  %%%%%%%%%%%%%%%%%%%%%%%%%%%%%%%%%%%%%%%%%%%%%
%%%%%%%%%%%%%%%%%%%%%%%%%%%%%%%%%%%%%%%%%%%%%%%%%%%%%%%%%%%%%%%%%%%%%%%%%%%%%%%%%%%%%%%

\newpage
\clearpage{\textbf{Figure 1}}
\vspace*{3cm}

\newpage
\clearpage{\textbf{Figure 2}}
\vspace*{4cm}

\newpage
\clearpage{\textbf{Figure 3}}
\vspace*{4cm}

\newpage
\clearpage{\textbf{Figure 4}}
\vspace*{4cm}

\newpage
\clearpage{\textbf{Figure 5}}
\vspace*{4cm}

\newpage
\clearpage{\textbf{Figure 6}}
\vspace*{0.4cm}

\newpage
\clearpage{\textbf{Figure 7}}
\vspace*{3cm}

\newpage
\clearpage{\textbf{Figure 8}}
\vspace*{3cm}

\newpage
\clearpage{\textbf{Figure 9}}
\vspace*{4cm}

\newpage
\clearpage{\textbf{Figure 10}}
\vspace*{4cm}

\newpage
\clearpage{\textbf{Figure 11}}
\vspace*{4cm}

\end{document}